# Attosecond Rabi Oscillations in High Harmonic Generation Resonantly Driven by Extreme Ultraviolet Laser Fields

Alba de las Heras,[1,2,*] Carlos Hernández-García,[1,2] Javier Serrano,[1,2] Aleksandar Prodanov,[3] Dimitar Popmintchev,[4] Tenio Popmintchev,[3,4] and Luis Plaja[1,2]

[1]*Grupo de Investigación en Aplicaciones del Láser y Fotónica,*
*Departamento de Física Aplicada, Universidad de Salamanca, Salamanca E-37008, Spain*
[2]*Unidad de Excelencia en Luz y Materia Estructuradas (LUMES), Universidad de Salamanca, Salamanca, Spain*
[3]*Department of Physics, Center for Advanced Nanoscience,*
*University of California San Diego, La Jolla, CA 92093, USA*
[4]*Photonic Institute, TU Wien, Vienna A-1040, Austria*
(Dated: March 27, 2024)

High-order harmonic generation driven by intense extreme ultraviolet (EUV) fields merges quantum optics and attosecond science, giving rise to an appealing route for the generation of coherent EUV and soft X-ray light for high-resolution imaging and spectroscopies. We theoretically investigate ultrafast resonant dynamics during the interaction of He atoms with strong extreme ultraviolet pulses. At high driving intensities, we identify record fast attosecond Rabi oscillations imprinting observable signatures in the high harmonic spectrum. At field strengths suppressing the Coulomb potential barrier for all the bounded states, we demonstrate the survival of the attosecond two-level dynamics for several Rabi cycles. Consequently, this intense EUV laser-atom interaction reveals a new strong-field scenario where the resonant coupling of two-level bound-bound transitions prevails, contrasting with the dominance of bound-continuum transitions in the conventional strong-field infrared regimes. These findings set an interesting perspective for extreme attosecond nonlinear optics with intense short-wavelength fields.

The advent of high intensity laser pulses in the extreme ultraviolet (EUV) region of the electromagnetic spectrum enables the exploration of nonlinear optics in the high frequency regime with unprecedented attosecond temporal resolution[1, 2]. Understanding attosecond extreme nonlinear optics is of interest for tracking fundamental electron-electron correlations[3–6], as well as for attosecond metrologies [7–9]. Series of breakthroughs in high-order harmonic generation (HHG) techniques to optimize the up-conversion efficiency [10–12], output power [13, 14] and peak intensity [15–18] provide the appealing fully spatially and temporarily coherent light to explore strong-field laser-matter interactions[19]. Remarkably, ultrafast quantum phenomena such as Rabi oscillations with periods as fast as 36 fs have been measured in the photoelectron dynamics of $Ar^+$ driven at EUV frequencies in a few-photon resonant interaction [20]. Also, bright EUV pulses from free-electron lasers have recently probed the resonance of autoionizing states[21], and the interference of few-photon pathways during Rabi cycles of 52 fs[22].

Among non-conventional attosecond EUV nonlinear optics phenomena, the physics of high harmonic up-conversion driven by an intense EUV pulse is yet to be fully investigated. Up to date, the standard approach for HHG in gases considers a lower frequency driving laser field at infrared, visible, or shorter ultraviolet wavelengths, reaching peak intensities around $10^{14}$-$10^{15}$ W/cm$^2$. In this range of parameters, the dominant mechanism of HHG is described by a three-step model [23, 24], involving i) tunnel ionization, ii) excursion of the electronic wavepacket in the continuum, and iii) recombination with the parent ion upon the conversion of the electron kinetic energy into the emission of high energy photons. Thus, in the low-frequency driving regime, HHG is founded in the strong coupling between the ground state and the continuum [25]. Longer wavelengths drive the electron further apart from the parent ion and imprint more kinetic energy. As a side-effect, the electron wavepacket diffusion during large excursions reduces the probability of recombination[25]. Consequently, mid-infrared lasers have been employed to prioritize high-photon energies[26], whereas near ultraviolet drivers enable higher brightness[10].

For EUV drivers, the physical picture of HHG differs significantly. The tunnel ionization probability is strongly suppressed compared to multiphoton ionization [27]. Additionally, resonant electron transitions are expected to play a relevant role, since the driving photon energies are in the order of photoexcitation energies. The few-states coupling with the laser field is characterized by a periodic electron population exchange known as Rabi oscillations[28]. Different works have unveiled femtosecond Rabi oscillations from the photoelectron spectra produced by an EUV[20, 22, 29] or near-infrared pulse[30], yet not under the conditions of full barrier suppression in the effective Coulomb potential. Rabi oscillations are relevant for identifying electron-electron correlations [31–34], for orbital angular momentum transfer from light to matter [35], for electromagnetic induced transparency [36], or for creating population gratings[37, 38]. In general, the physics of coherently driven atomic resonances itself is of a paramount research interest[39, 40].

In principle, the period of Rabi oscillations can be shortened to attosecond timescales by using very intense EUV pulses. Still, an intriguing scenario arises when the electric field amplitude is large enough to deform the atomic potential leading to a transient suppression of the Coulomb barrier for all the bound states[41, 42]. In this nontrivial situation, the rapid depletion of the bound electron population should affect the resonant transition, eventually dismantling the conventional coherent two-level dynamics.

In this letter, we theoretically investigate the resonant electron wavepacket dynamics in He atoms interacting with strong EUV laser pulses. We demonstrate the generation of high order harmonics reaching the soft X-rays, while encoding characteristic signatures of ultrafast Rabi oscillations both in the spectral and temporal domains. We find that this extreme up-conversion is dominated by ultrafast two-state dynamics, even in the case of strong barrier suppression of all the atomic levels. Surprisingly, the two-level dynamics prevails even when the two energy levels are above the potential barrier. The perseverance of the two-level dynamics under such strong ionization conditions can be explained by considering a two-level atomic model that includes uneven level-ionization rates. As described in ref. [43], the resonant interaction has a stabilizing effect, since the field-dressed states are found to ionize slower than the fastest rate of the atomic levels[44]. This coherent dressing is critical for the persistence of attosecond Rabi oscillations, since it prevents the rapid depletion of the excited state at the high driving intensities and high driving frequencies required for Rabi periods in the attosecond timescale. Unlike previous mechanisms of ionization suppression[45–50], this stabilization behavior appears exclusively in the vicinity of a resonance. In addition, our exact simulations show a substantial return of the continuum electron population to the ground state after the interaction reaches its maximum. An *ad hoc* modification of the level ionization rates to include this recombination in our two-level atomic model shows this as a second reason for the resilience of the two-level dynamics[43].

Our exact calculations correspond to the *ab initio* numerical integration of the two-electron dynamics in a He atom along the laser polarization dimension[51]. In the numerical integration of the 1D time-dependent Schrödinger equation, we consider two active correlated electrons. Our calculations include the exact description of both neutral and ionic species, He and He$^+$, during the ultrafast laser interaction. Here, the Hamiltonian in atomic units (a.u.) is given by:

$$\hat{H} = \sum_{j=1}^{2} \left\{ \frac{1}{2}\left(\hat{p}_{x,j} + \frac{A(t)}{c}\right)^2 + V_j \right\} + V_{1,2}, \quad (1)$$

where the Coulomb potentials, $V_j(x_j) = -\frac{2}{\sqrt{x_j^2 + a_N}}$ and $V_{1,2}(x_1, x_2) = \frac{1}{\sqrt{(x_1-x_2)^2 + a_{ee}}}$, include the softening parameters $a_{ee} = 0.32$ a.u. and $a_N = 0.50$ a.u. to match the actual ionization energies, $I_p^{He} = 24.5$ eV and $I_p^{He^+} = 54.4$ eV. $\hat{p}_{x,j} = -i\frac{\partial}{\partial x_j}$ is the momentum operator, $c$ is the speed of light in vacuum, and $A(t)$ is the vector potential related to the electric field by $E(t) = -\frac{1}{c}\frac{\partial A(t)}{\partial t}$. The initial ground state is propagated via the finite-difference Crank-Nicholson scheme with spatial and temporal steps of $\delta x = 0.2$ a.u., and $\delta t = 0.05$ a.u. The stationary atomic states are found from the field-free Hamiltonian diagonalization.

We compute the dipole acceleration by applying the Ehrenfest theorem:

$$\langle a(t) \rangle = -\langle \Psi(t) | \nabla V(x_1, x_2) | \Psi(t) \rangle, \quad (2)$$

being $V(x_1, x_2) = V_1(x_1) + V_2(x_2) + V_{1,2}(x_1, x_2)$, and $\nabla = \left(\frac{\partial}{\partial x_1} + \frac{\partial}{\partial x_2}\right)$. Eq. 2 accounts for the total dipole emission from He and He$^+$, including the contribution from other pathways different from the two-level dynamics. To isolate the emission from the two-level dipole transition, we calculate the partial component of the acceleration

$$\langle a_{1s^2, 1s2p}(t) \rangle = -c_{1s^2}^*(t) c_{1s2p}(t) \langle 1s^2 | \nabla V(x_1, x_2) | 1s2p \rangle, \quad (3)$$

where $c_{1s^2}^*(t) = \langle \Psi(t) | 1s^2 \rangle$, and $c_{1s2p}(t) = \langle 1s2p | \Psi(t) \rangle$. The harmonic spectra are computed as the Fourier transform of the dipole acceleration, using Eq. (2) to obtain the total emission or Eq. (3) to select the contribution from the two-level dipole transition.

The electric field is described as a sinusoidal squared envelope of 32-cycles at the full pulse duration, $E(t) = E_0 \sin^2(t\omega_0/64) \sin(\omega_0 t)$. We choose driving photon energies of 12.4–24.4 eV around the atomic resonance $1s^2$–$1s2p$. The peak intensities lie in the range $8.8 \times 10^{13}$–$8.8 \times 10^{15}$ W/cm$^2$ (corresponding to field amplitudes of $E_0 = 0.05$–$0.5$ a.u.). This implies values of the Keldysh parameter[27] of 1.8–18.1, associated with multiphoton ionization, and corresponding to extremely short electron excursions[24] of less than 1 a.u.

The effective Coulomb potential $V_1(x) + V_{1,2}(x, 0)$ is depicted in Fig. 1a at a peak intensity of $1.6 \times 10^{15}$ W/cm$^2$ (peak field amplitude of $E_0 = 0.2154$ a.u.), where the energy levels from which the population of the resonant states ionizes, $-I_p^{(1s^2)}$ and $-I_p^{(1s2p)}$, lie above the suppressed potential barrier. Note that the potential is plotted in terms of the position of one of the electrons, $x_1$, while the other remains frozen at its mean position, $x_2 = 0$. It is worth pointing out that the excited level surpasses the potential barrier at lower field amplitudes than the ground state. Consequently, the excited state lies in the continuum for a longer time during the interaction with the electric EUV pulse, leading to a greater ionization compared to the ground state. This justifies the uneven ionization rates used in our two-level model presented in ref.[43]. The central photon energy of the driving electric field (Fig. 1b) is chosen to match the atomic $1s^2 - 1s2p$ transition at 18.4 eV, corresponding to a central wavelength of 67.5 nm or a driving frequency of $\omega_0 = 0.676$ a.u.

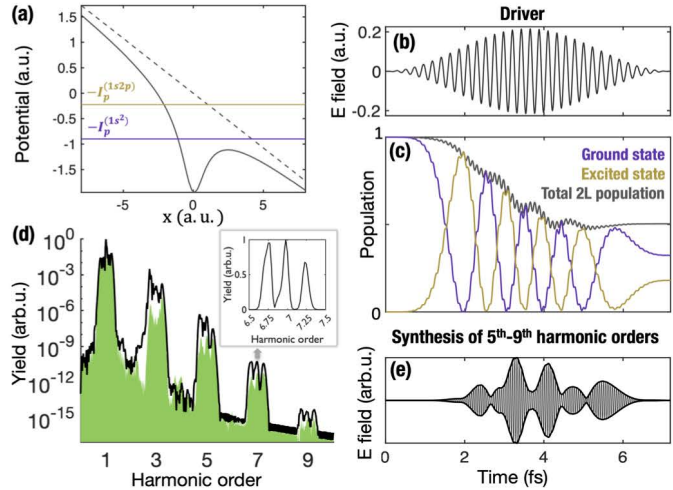

FIG. 1: (a) Effective Coulomb potential distorted by the intense EUV driving pulse, and energy levels from which the population of the $1s^2$ and $1s2p$ states ionizes: $-I_p^{(1s^2)}$ and $-I_p^{(1s2p)}$. (b) Femtosecond EUV driving electric pulse. (c) Populations of the ground state (purple) and the excited state (gold) reveal Rabi oscillations with a sub-femtosecond period of $\sim 854$ as. The total two-level population (gray) shows a step-decay following the Rabi cycles and a certain stabilization in the second part of the interaction. (d) Corresponding high harmonic spectrum which encodes the attosecond Rabi dynamics as a characteristic three-peak structure. The black solid line is the spectrum of the total acceleration, Eq. (2), whereas the green-filled area corresponds to the contribution from the two-level dipole transition, Eq. (3). The inset shows the three-peak structure of the $7^{th}$ harmonic on linear scale. (e) The synthesis of the high-order harmonics (from $5^{th}$ to $9^{th}$) results in a pulse modulation on an attosecond timescale.

The population of each atomic level, $1s^2$ and $1s2p$, shown in Fig. 1c, reveals unprecedentedly rapid attosecond Rabi oscillations. The Rabi periodic modulations of $\sim 854$ as extend over multiple Rabi cycles since the depletion of the two-level transition is on a much slower timescale of several femtoseconds. The temporal evolution of the total population occupying the atomic levels $1s^2$ and $1s2p$ is also shown in Fig. 1c. Remarkably, half of the initial population remains in these two levels at the end of the laser pulse, despite the Coulomb barrier suppression conditions. Besides, the total two-state population shows a characteristic step-like decay that follows the Rabi flopping, instead of the laser period dictating the typical strong field interaction with low-frequency infrared drivers. This suggests that the relevant modulation of the ionization is not connected with the transient suppression of the Coulomb barrier which induces a faster modulation, also observable in the calculations. Rather, the step-like dynamics are correlated with the coherent dynamics of the two-level transition. Additionally, the ionization rate is significantly reduced after the peak amplitude of the driving field is reached. All these important dynamics which are significantly different compared to the low-frequency driving regimes are further addressed in ref.[43].

In the frequency domain, the high-frequency Rabi oscillations give rise to a distinct signature in the harmonic spectrum (black solid line in Fig. 1d), therefore allowing for experimental characterization. Each harmonic order, $q$, splits into a triplet structure, where we identify the usual central peak and two additional satellite peaks at frequencies $q\omega_0 - \Omega$ and $q\omega_0 + \Omega$. Here, the Rabi frequency is $\Omega \approx 0.263\,\omega_0$. The triplet structure is known to be replicated in HHG in strongly driven two-level dynamics[52, 53]. In the EUV strong-field regime, the multi-peak trace exhibits three well-resolved peaks of similar intensity (see the $7^{th}$ harmonic on linear scale in the inset).

In addition to the analysis of the total dipole emission, the green-filled curve in Fig. 1d shows the harmonic spectrum associated with the two-level transition dipole, calculated using Eq. (3). We identify the two-state coupling as an important contribution to the high harmonic spectrum, exhibiting also the multi-peak structure. The fact that this emission does not reproduce the full high harmonic spectrum suggests that the coupling of the resonant states with the energy continuum is also relevant in EUV-driven high harmonic generation, as expected in this high-intensity regime[44].

Interestingly, in the time domain, the synthesis of the emission from the higher-order harmonics (from 5th to 9th) yields a femtosecond pulse with an envelope that is modulated on the attosecond scale due to the Rabi flopping (Fig. 1e). The yield of a two-level dipole emission is minimized when either of the two states is depleted. This introduces a prominent difference compared to the well-structured attosecond pulse trains with a half-laser-cycle period, emerging in conventional HHG from low-frequency infrared drivers[54]. This further illustrates the peculiarities of this strong-field EUV regime.

Up to this point, we have described the resonant atomic behavior. The dependence on the driving photon energy is shown in Figs. 2a,b, whilst the peak intensity is maintained at $1.6 \times 10^{15}$ W/cm$^2$. We observe attosecond Rabi oscillations in the populations of the ground (Fig. 2a) and excited (Fig. 2b) states in the vicinity of the atomic resonance (indicated with a green dotted line). The frequency of the Rabi oscillations increases with the detuning between the driving field frequency ($\omega_0$) and the frequency of the atomic transition ($\omega_a$), $\Delta = \omega_a - \omega_0$. This is in agreement with the equation of the generalized Rabi frequency in a two-level system[55]:

$\Omega = \sqrt{\Delta^2 + |\chi|^2}$, where $|\chi|^2$ is the intensity of the coupling, and the longest Rabi period corresponds to the resonant case $\Delta = 0$. Still, a slight energy shift of the resonant transition ($\sim 1$ eV) arises during the interaction with the laser pulse, most probably resulting from the field dressing.

We also note that the total two-state population remarkably depends on the driving photon energy (Figs. 2a,b). There is a strong depletion far away from the resonance, whereas the two-level population exhibits stabilized dynamics for small detunings. Intriguingly, the behavior is not symmetric for positive and negative detunings. At photon energies slightly below the resonance, the population is more stable. By introducing a certain detuning, the Rabi oscillations become less efficient in the population transfer to the excited state —which has a higher ionization rate $\gamma_2(t) > \gamma_1(t)$—, so one would expect a lower depletion of the total two-level population, as observed for small positive detunings. In contrast, at slightly higher energies above the resonance, we notice narrow channels of strong and low depletion that might be attributed to other atomic resonances.

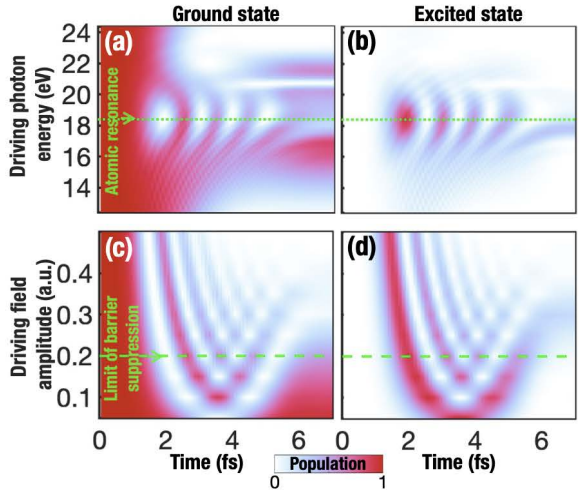

FIG. 2: Ultrafast evolution of the populations of (a) the ground state and (b) the excited state of He for different driving photon energies, at a peak intensity of $1.6 \times 10^{15}$ W/cm$^2$, illustrating femtosecond-to-attosecond Rabi oscillations in a broad range of energies around the resonance. Panels (c) and (d) show the direct increase of the frequency of the Rabi oscillations to attosecond timescales (to sub-500 as) with the driving field amplitude for the resonant case of 18.4 eV.

The scaling of the evolution of the population with the driving field amplitude (Figs. 2c,d) also deserves special attention. Increasing the field amplitude implies more ionization and faster Rabi oscillations. The two effects are expected since i) we are enhancing the bound-continuum coupling, and ii) the Rabi frequency is directly proportional to the amplitude of the resonant driver[55]. Note the pronounced ionization only occurs well beyond the limit of barrier suppression for all bounded states, which again suggests that the reduction of the ionization rate is due to resonant stabilization. This fact is of paramount importance for the persistence of Rabi oscillations with periods even shorter than 500 as.

Moving once again to the frequency domain, the high harmonic spectrum in Fig. 3a presents a map of the satellite-peak structure around the resonance (also shown in Fig. 1d). Identically to the two-level picture, we recover the single peak far off of the resonance. This wide range of detunings at which Rabi oscillations can be observed should ease the experimental traceability of these two-level coherent dynamics in He in the strong high-frequency interaction regime.

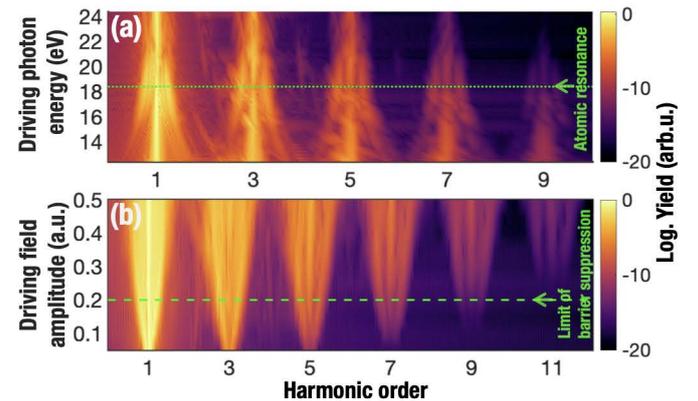

FIG. 3: (a) High harmonic spectrum as a function of the driving photon energy at a fixed peak intensity of $1.6 \times 10^{15}$ W/cm$^2$, showing the well-pronounced three-peak spectroscopic signature of the attosecond Rabi oscillations at resonance. (b) High harmonic spectrum as a function of the driving field amplitude at the resonant photon energy of 18.4 eV.

As the EUV field amplitude increases (see Fig. 3b), the satellite peaks in the high harmonic spectrum separate, also as expected for a two-level system. Despite the depletion of the bound states, the HHG efficiency rises for stronger field amplitudes and the multi-peak signature is preserved. The harmonic signal is observed up to the $11^{th}$ harmonic, corresponding to a soft X-ray photon energy of 202 eV. A closer inspection of Fig. 3b indicates that the relevance of the central harmonic peak increases at higher driving intensities. This suggests that its origin is associated with the bound-continuum coupling, instead of the two-level transition (as further corroborated in ref. [43]).

In summary, we have studied the generation of high order harmonics using an EUV driver both as an alternative soft X-ray coherent source and as a powerful spectroscopic tool. Our *ab initio* calculations reveal attosecond Rabi oscillations in the resonant dynamics of EUV-driven HHG. To the best of our knowledge, these are the fastest Rabi oscillations ever reported. Despite the high EUV intensities required to reach Rabi periods in the attosecond scale, the coupled two-state behavior survives during the strong field interaction. Using a simplified two-level model including uneven level-ionization rates, we demonstrate that the dressed states deplete at a slower ionization rate than the fastest rate of the bare states, stabilizing the coupled two-state population. The signature of the ultrafast Rabi oscillations is uniquely encoded in the EUV-driven high harmonic spectrum as a multi-peak harmonic structure, allowing for experimental characterization. We envision the EUV strong-field interactions as a new fascinating area of extreme nonlinear optics and attosecond quantum optics radically different from its lower-frequency counterpart, now accessible with cutting-edge high harmonic sources and EUV free-electron lasers.

We acknowledge funding from the European Research Council (ERC) under the European Union's Horizon 2020 research and innovation programme (grant agreement No 851201), as well as from the projects PID2019-106910GB-I00 and PID2022-142340NB-I00 financed by Ministerio de Ciencia e Innovacción and Agencia Estatal de Investigación (MCIN/ AEI/10.13039/501100011033/). TP acknowledges funding from the European Research Council (ERC) under the European Union's Horizon 2020 research and innovation program (grant agreement XSTREAM-716950), and from the Alfred P. Sloan Foundation (FG-2018-10892).

---

* albadelasheras@usal.es
[1] G. Sansone, L. Poletto, and M. Nisoli, Nature Photonics **5**, 655 (2011).
[2] J. Duris, S. Li, T. Driver, E. G. Champenois, J. P. MacArthur, A. A. Lutman, Z. Zhang, P. Rosenberger, J. W. Aldrich, R. Coffee, et al., Nature Photonics **14**, 30 (2020).
[3] Y. Nabekawa, H. Hasegawa, E. J. Takahashi, and K. Midorikawa, Physical Review Letters **94**, 1 (2005).
[4] K. Midorikawa, Y. Nabekawa, and A. Suda, Progress in Quantum Electronics **32**, 43 (2008).
[5] P. Tzallas, E. Skantzakis, L. A. A. Nikolopoulos, G. D. Tsakiris, and D. Charalambidis, Nature Physics **7**, 781 (2011).
[6] M. Kretschmar, A. Hadjipittas, B. Major, J. Tümmler, I. Will, T. Nagy, M. J. J. Vrakking, A. Emmanouilidou, and B. Schütte, Optica **9**, 639 (2022).
[7] P. Tzallas, D. Charalambidis, N. A. Papadogiannis, K. Witte, and G. D. Tsakiris, Nature **426**, 267 (2003).
[8] T. Sekikawa, A. Kosuge, T. Kanai, and S. Watanabe, Nature **432**, 605 (2004).
[9] I. Orfanos, I. Makos, I. Liontos, E. Skantzakis, B. Förg, D. Charalambidis, and P. Tzallas, APL Photonics **4**, 080901 (2019).
[10] D. Popmintchev, C. Hernandez-Garcia, F. Dollar, C. Mancuso, J. A. Perez-Hernandez, M.-C. Chen, A. Hankla, X. Gao, B. Shim, A. L. Gaeta, et al., Science **350**, 1225 (2015).
[11] H. Wang, Y. Xu, S. Ulonska, J. S. Robinson, P. Ranitovic, and R. A. Kaindl, Nature Communications **6**, 1 (2015).
[12] F. Ferrari, F. Calegari, M. Lucchini, C. Vozzi, S. Stagira, G. Sansone, and M. Nisoli, Nature Photonics **4**, 875 (2010).
[13] E. J. Takahashi, P. Lan, O. D. Mücke, Y. Nabekawa, and K. Midorikawa, Nature Communications **4**, 2691 (2013).
[14] B. Xue, Y. Tamaru, Y. Fu, H. Yuan, P. Lan, O. D. Mücke, A. Suda, K. Midorikawa, and E. J. Takahashi, Science Advances **6** (2020).
[15] H. Mashiko, A. Suda, and K. Midorikawa, Optics Letters **29**, 1927 (2004).
[16] B. Bergues, D. E. Rivas, M. Weidman, A. A. Muschet, W. Helml, A. Guggenmos, V. Pervak, U. Kleineberg, G. Marcus, R. Kienberger, et al., Optica **5**, 237 (2018).
[17] B. Major, O. Ghafur, K. Kovács, K. Varjú, V. Tosa, M. J. J. Vrakking, and B. Schütte, Optica **8**, 960 (2021).
[18] B. Senfftleben, M. Kretschmar, A. Hoffmann, M. Sauppe, J. Tümmler, I. Will, T. Nagy, M. J. Vrakking, D. Rupp, and B. Schütte, Journal of Physics: Photonics **2**, 034001 (2020).
[19] X. Shi, C.-T. Liao, Z. Tao, E. Cating-Subramanian, M. M. Murnane, C. Hernández-García, and H. C. Kapteyn, Journal of Physics B: Atomic, Molecular and Optical Physics **53**, 184008 (2020).
[20] M. Flögel, J. Durá, B. Schütte, M. Ivanov, A. Rouzée, and M. J. Vrakking, Physical Review A **95**, 021401 (2017).
[21] C. Ott, L. Aufleger, T. Ding, M. Rebholz, A. Magunia, M. Hartmann, V. Stooß, D. Wachs, P. Birk, G. D. Borisova, et al., Physical Review Letters **123**, 163201 (2019).
[22] S. Nandi, E. Olofsson, M. Bertolino, S. Carlström, F. Zapata, D. Busto, C. Callegari, M. Di Fraia, P. Eng-Johnsson, R. Feifel, et al., Nature **608**, 488 (2022).
[23] K. J. Schafer, B. Yang, L. F. DiMauro, and K. C. Kulander, Phys. Rev. Lett. **70**, 1599 (1993).
[24] P. B. Corkum, Phys. Rev. Lett. **71**, 1994 (1993).
[25] M. Lewenstein, P. Balcou, M. Y. Ivanov, A. L'Huillier, and P. B. Corkum, Physical Review A **49**, 2117 (1994).
[26] T. Popmintchev, M. C. Chen, D. Popmintchev, P. Arpin, S. Brown, S. Ališauskas, G. Andriukaitis, T. Balčiunas, O. D. Mücke, A. Pugzlys, et al., Science **336**, 1287 (2012).
[27] L. V. Keldysh, Journal of Experimental and Theoretical Physics **20**, 1307 (1965).
[28] I. I. Rabi, Physical Review **49**, 324 (1936).
[29] W. C. Jiang, H. Liang, S. Wang, L. Y. Peng, and J. Burgdörfer, Physical Review Research **3** (2021).
[30] M. Fushitani, C.-N. Liu, A. Matsuda, T. Endo, Y. Toida, M. Nagasono, T. Togashi, M. Yabashi, T. Ishikawa, Y. Hikosaka, et al., Nature Photonics **10**, 102 (2016).
[31] Q. Liao, Y. Zhou, C. Huang, and P. Lu, New Journal of Physics **14**, 013001 (2012), ISSN 1367-2630, URL https://iopscience.iop.org/article/10.1088/1367-2630/14/1/013001.
[32] Y. O. Dudin, L. Li, F. Bariani, and A. Kuzmich, Nature Physics **8**, 790 (2012).
[33] U. De Giovannini, G. Brunetto, A. Castro, J. Walkenhorst, and A. Rubio, ChemPhysChem **14**, 1363 (2013).
[34] Y. Chen, Y. Zhou, Y. Li, M. Li, P. Lan, and P. Lu, Physical Review A **97** (2018).
[35] C. T. Schmiegelow, J. Schulz, H. Kaufmann, T. Ruster, U. G. Poschinger, and F. Schmidt-Kaler, Nature Com-

munications **7**, 1 (2016).
[36] P. Ranitovic, X. M. Tong, C. W. Hogle, X. Zhou, Y. Liu, N. Toshima, M. M. Murnane, and H. C. Kapteyn, Physical Review Letters **106**, 193008 (2011).
[37] R. M. Arkhipov, Optics and Spectroscopy **128**, 1865 (2020).
[38] R. Arkhipov, A. Pakhomov, M. Arkhipov, I. Babushkin, A. Demircan, U. Morgner, and N. Rosanov, Scientific Reports **11**, 1961 (2021).
[39] V. Strelkov, Physical Review Letters **104**, 123901 (2010).
[40] A. Kaldun, A. Blättermann, V. Stooß, S. Donsa, H. Wei, R. Pazourek, S. Nagele, C. Ott, C. D. Lin, J. Burgdörfer, et al., Science **354**, 738 (2016).
[41] S. Augst, D. Strickland, D. D. Meyerhofer, S. L. Chin, and J. H. Eberly, Physical Review Letters **63**, 2212 (1989).
[42] S. Augst, D. D. Meyerhofer, D. Strickland, and S. L. Chint, Journal of the Optical Society of America B **8**, 858 (1991).
[43] See Supplemental Material at [URL will be inserted by publisher] for the description of the two-level modeling and additional results.
[44] S. V. Popruzhenko, V. D. Mur, V. S. Popov, and D. Bauer, Physical Review Letters **101**, 193003 (2008), ISSN 0031-9007, 0807.2226, URL https://link.aps.org/doi/10.1103/PhysRevLett.101.193003.
[45] R. J. Vos and M. Gavrila, Physical Review Letters **68**, 170 (1992).
[46] J. H. Eberly and K. C. Kulander, Science **262**, 1229 (1993).
[47] I. P. Christov, J. Zhou, J. Peatross, A. Rundquist, M. M. Murnane, and H. C. Kapteyn, Physical Review Letters **77**, 1743 (1996).
[48] A. M. Popov, O. V. Tikhonova, and E. A. Volkova, Journal of Physics B: Atomic, Molecular and Optical Physics **36**, R125 (2003).
[49] U. Eichmann, A. Saenz, S. Eilzer, T. Nubbemeyer, and W. Sandner, Physical Review Letters **110**, 203002 (2013), ISSN 0031-9007.
[50] Q. Wei, P. Wang, S. Kais, and D. Herschbach, Chemical Physics Letters **683**, 240 (2017).
[51] A. de las Heras, C. Hernández-García, and L. Plaja, Physical Review Research **2**, 033047 (2020).
[52] L. Plaja and L. Roso-Franco, Journal of the Optical Society of America B **9**, 2210 (1992).
[53] A. Picón, L. Roso, J. Mompart, O. Varela, V. Ahufinger, R. Corbalán, and L. Plaja, Physical Review A **81**, 033420 (2010).
[54] P. M. Paul, E. S. Toma, P. Breger, G. Mullot, F. Augé, P. Balcou, H. G. Muller, and P. Agostini, Science **292**, 1689 (2001).
[55] C. J. Joachain, N. J. Kylstra, and R. M. Potvliege, *Atoms in Intense Laser Fields* (Cambridge University Press, 2011).

# Supplementary Materials:
# Attosecond Rabi Oscillations in High Harmonic Generation Resonantly Driven by Extreme Ultraviolet Laser Fields

Alba de las Heras,[1,2,*] Carlos Hernández-García,[1,2] Javier Serrano,[1,2] Aleksandar Prodanov,[3] Dimitar Popmintchev,[4] Tenio Popmintchev,[3,4] and Luis Plaja[1,2]

[1]*Grupo de Investigación en Aplicaciones del Láser y Fotónica,*
*Departamento de Física Aplicada, Universidad de Salamanca, Salamanca E-37008, Spain*
[2]*Unidad de Excelencia en Luz y Materia Estructuradas (LUMES),*
*Universidad de Salamanca, Salamanca, Spain*
[3]*Department of Physics, Center for Advanced Nanoscience,*
*University of California San Diego, La Jolla, CA 92093, USA*
[4]*Photonic Institute, TU Wien, Vienna A-1040, Austria*

* albadelasheras@usal.es

These supplementary materials cover the two-level modeling using uneven ionization rates which helps to describe intuitively the stabilization of the two-state dynamics in He and provides additional data to support the core findings presented in the main manuscript. We begin with a description of the two-level model and the comparison of these results with the full-*ab initio* calculation. Then, in the second section, we use the Rotating-Wave-Approximated Hamiltonian to analyze the dressed-state two-level dynamics, underpinning analytical equations describing the role of the ionization rates asymmetry.

## I. TWO-LEVEL MODEL WITH UNEVEN IONIZATION RATES

We consider the equations of the dynamics in a two-level system featuring uneven damping functions $\gamma_1(t)$ and $\gamma_2(t)$:

$$\dot{c}_1(t) = -i\frac{1}{2}\left[-(\omega_a + i\gamma_1(t))c_1(t) - \left(\chi(t)e^{-i\omega_0 t} + \chi^*(t)e^{i\omega_0 t}\right)c_2(t)\right] \quad (1)$$

$$\dot{c}_2(t) = -i\frac{1}{2}\left[(\omega_a - i\gamma_2(t))c_2(t) - \left(\chi(t)e^{-i\omega_0 t} + \chi^*(t)e^{+i\omega_0 t}\right)c_1(t)\right] \quad (2)$$

Here, the driving laser frequency is denoted as $\omega_0$, and the atomic transition frequency as $\omega_a$. The coupling strength is modeled as $\chi(t) = \chi_0 \operatorname{Env}(t)$, with $\chi_0 = 0.178$ a.u. being the peak coupling strength, and $\operatorname{Env}(t) = \sin^2(t\omega_0/64)$ the driving pulse envelope of 32 cycles. The numerical solution of Eqs.(1) and (2) is found using a standard Runge-Kutta method.

First, we estimate $\gamma_1(t)$ and $\gamma_2(t)$ using the model in Ref. [1] to mimic the standard ionization from each level (see Fig. 1a). Figure 1b shows the comparison of the population evolution in the two-level model (2L model in dashed lines) and the accurate full time-dependent Schrödinger equation (full TDSE in solid lines). While this simplified 2L theoretical description perfectly reproduces the frequency of the Rabi oscillations obtained with the TDSE model, the the population decay is overestimated.

As an alternative method to evaluate the damping functions, we perform best-fit analyses (2L-BF) to match the full TDSE calculations by tuning the damping functions $\gamma_{1,BF}(t)$ and $\gamma_{2,BF}(t)$ (see Fig. 1c). A key element here is to include the return of the population to the ground state, i.e. $\gamma_{1,BF}(t)$ flips its sign in the second part of the interaction. Thus, in this damping model, we consider both the ionization rate, as well as the recombination mechanism that increases the ground state population later in the interaction. The excellent agreement of this 2L-BF best-fit two-state dynamics to the exact TDSE calculations is show

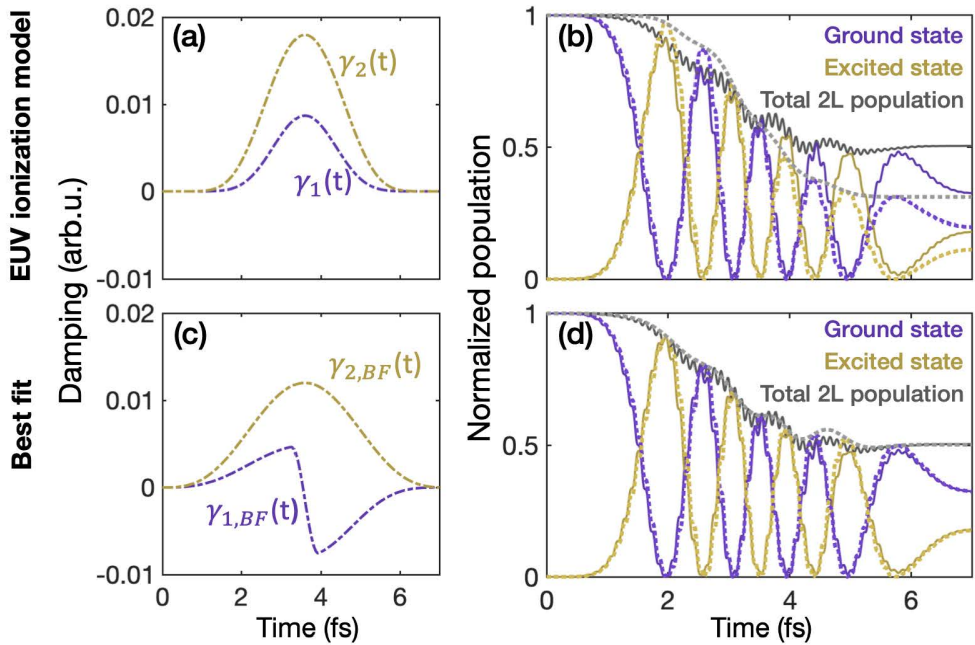

FIG. 1. (a) Ionization rates of the ground (purple) and excited (gold) state using the theory of Ref. [1]. (b) Comparison of the evolution of the population of the states obtained using the two-level model (dashed lines) versus the *ab initio* TDSE description (solid lines), showing identical periods of Rabi oscillation, while the ionization rates are overestimated by the two-level model. (c) and (d) Comparison of the equivalent plots using a best-fit two-level model versus the exact TDSE solution showing an excellent agreement in ionization rates. We consider a resonant EUV driving pulse with a central photon energy of $\omega_0 = \omega_a = 18.4$ eV and a peak intensity of $1.6 \times 10^{15}$ W/cm$^2$.

in Fig. 1d.

Finally, we compute the 2L dipole spectrum as $\Re\{c_1(t)c_2^*(t)\}$. The high harmonic spectrum from the *ab initio* TDSE calculation is shown in black in Fig. 2, whereas the blue area corresponds to the 2L model with the standard ionization rates $\gamma_1(t)$ and $\gamma_2(t)$. Note that the reduced population of the bound states does not lead to a significant change in the spectral shape. Despite the lower yield resulting in the 2L model, the position of the satellite peaks is perfectly matched. The much lower yield of the higher order harmonics (7th, 9th) suggests that the high frequency dipole oscillations are amplified by the coupling with the continuum. Furthermore, the absent central peak in the 2L model is also an indication of the relevance of the coupling to the continuum, as emphasized in the main text.

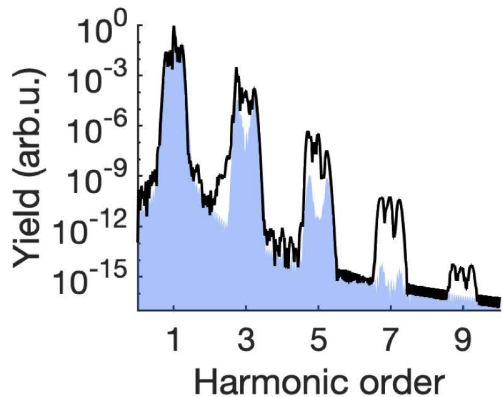

FIG. 2. Comparison of EUV-driven high harmonic spectra obtained using the 2L two-level model (blue area), compared to the exact TDSE calculation (black line). The strong coupling to the continuum states for the higher order harmonics (5th, 7th) is indicated by their reduced intensity and the absent central peak.

## II. DRESSED EIGENVALUES IN THE ROTATING WAVE APPROXIMATION

The ionization rates presented in Figs. 1a and c have a pronounced asymmetry ($\gamma_2 \gg \gamma_1$). Naïvely, it might seem that the average lifetime of the two-level transition would be predominantly influenced by the quickest ionization process, namely, the ionization of the excited state, characterized by a mean ionization time $\overline{\tau}_2 = 1/\overline{\gamma}_2 \sim 4.8$ fs. Figures 1b and c demonstrate that this is not the cas, as the Rabi oscillations extend over a time much larger than $\overline{\tau}_2$. We resort to the dressed-state picture to understand this robust stabilization dynamics of the two-level transition with respect to ionization.

The energy of the dressed levels can be computed by diagonalizing the rotating-wave approximated (RWA) Hamiltonian:

$$H_{RWA} = \frac{1}{2} \begin{pmatrix} -\Delta'_1 & -\chi^* \\ -\chi & \Delta'^*_2 \end{pmatrix}, \qquad (3)$$

which is non-hermitian due to the inclusion of the ionization dynamics. The complex detunings are defined as $\Delta'_1 = \Delta + i\gamma_1$ and $\Delta'_2 = \Delta + i\gamma_2$, where $\Delta = \omega_a - \omega_0$. The Hamiltonian diagonalization leads to two complex-valued dressed state energies $\lambda_-(t)$ and $\lambda_+(t)$:

$$\lambda_\pm(t) = \frac{\overline{\gamma}(t)}{2} \left\{ -i \pm \left[ \left( \frac{\Delta}{\overline{\gamma}(t)} - i\alpha(t) \right)^2 + \frac{|\chi(t)|^2}{\overline{\gamma}(t)^2} \right]^{1/2} \right\} \qquad (4)$$

where $\bar{\gamma}(t) = (\gamma_1(t) + \gamma_2(t))/2$, and the $\gamma$-asymmetry factor is defined as $\alpha(t) = (\gamma_2(t) - \gamma_1(t))/(\gamma_2(t) + \gamma_1(t))$. Considering the parameter space of our EUV strong-field interaction, the first term is the dominant one. Hence, the population decay of the dressed eigenstates is essentially dominated by $\bar{\gamma}$. The $\gamma$-asymmetry explains the step-like damping of the Rabi oscillations in Fig. 1d, both in the ab-initio TDSE and 2L-BF descriptions. A more substantial depletion occurs at the Rabi half periods where the population is mainly in the excited state, contrasting with the lower depletion when the ground state is highly populated. Consequently, the total 2L population which is periodically exchanged between the two states depletes with $\bar{\gamma}(t)$.

The generalized Rabi frequency can be expressed as:

$$\Omega = \Re\{\lambda_+ - \lambda_-\} = \Re\left\{\sqrt{(\Delta - i\alpha(t)\bar{\gamma}(t))^2 + |\chi(t)|^2}\right\}, \tag{5}$$

which in the case of even damping, i. e. $\gamma_1(t) = \gamma_2(t)$, reduces to:

$$\Omega = \sqrt{\Delta^2 + |\chi|^2}. \tag{6}$$

---

[1] S. V. Popruzhenko, V. D. Mur, V. S. Popov, and D. Bauer, Physical Review Letters **101**, 193003 (2008), arXiv:0807.2226.